\documentstyle[12pt]{article}
\renewcommand{\baselinestretch}{1.5}

\newtheorem{Thm}{Theorem}

\newtheorem{Lem}[Thm]{Lemma}
\newtheorem{Prop}[Thm]{Proposition}

\newenvironment{proof}{\noindent {\bf proof: }}{\QED}

\newcounter{masection}
\newcommand{\masection}[1]{\setcounter{equation}{0}
    \refstepcounter{masection} \vspace{10pt plus 10pt minus 3pt}
    \noindent
    {\large\bf \arabic{masection} #1}\par\vspace{5pt}}

\renewcommand{\theequation}{\mbox{\arabic{masection}.\arabic{equation}}}

\newcounter{masubsection}[masection]
\newcommand{\masubsection}[1]{
    \refstepcounter{masubsection} \vspace{5pt plus 5pt minus 2pt}
    \renewcommand{\themasubsection}{%
        \arabic{masection}.\arabic{masubsection}}
    \noindent
    {\bf \arabic{masection}.\arabic{masubsection} #1}
    \par\vspace{5pt}}

\newcounter{saveeqn}
\newcommand{\alpheqn}{\setcounter{saveeqn}{\value{equation}}%
  \stepcounter{saveeqn}\setcounter{equation}{0}%
  \renewcommand{\theequation}{%
  \mbox{\arabic{masection}.\arabic{saveeqn}-\alph{equation}}}}
\newcommand{\reseteqn}{\setcounter{equation}{\value{saveeqn}}%
\renewcommand{\theequation}{\mbox{%
    \arabic{masection}.\arabic{equation}}}}
\newcommand{\eq}[1]{(\ref{#1})}

\newcommand{\be}{\begin{equation}}
\newcommand{\ee}{\end{equation}}
\newcommand{\bea}{\begin{eqnarray}}
\newcommand{\eea}{\end{eqnarray}}
\newcommand{\beann}{\begin{eqnarray*}}
\newcommand{\eeann}{\end{eqnarray*}}

\newcommand{\Hs}{{\cal H}}

\newcommand{\up}{\uparrow}
\newcommand{\down}{\downarrow}

\newcommand{\ket}[1]{\left\vert #1\right\rangle}

\def\idty{{\leavevmode{\rm 1\ifmmode\mkern -5.4mu\else
    \kern -.3em\fi I}}}
\def\Ibb #1{ {\rm I\ifmmode\mkern -3.6mu\else\kern -.2em\fi#1}}
\def\Ird{{\hbox{\kern2pt\vbox{\hrule height0pt depth.4pt width5.7pt
    \hbox{\kern-1pt\sevensy\char"36\kern2pt\char"36} \vskip-.2pt
    \hrule height.4pt depth0pt width6pt}}}}
\def\Irs{{\hbox{\kern2pt\vbox{\hrule height0pt depth.34pt width5pt
    \hbox{\kern-1pt\fivesy\char"36\kern1.6pt\char"36} \vskip -.1pt
    \hrule height .34 pt depth 0pt width 5.1 pt}}}}
\def\Ir{\mbox{\bf Z}}
\def\ibb #1{\leavevmode\hbox{\kern.3em\vrule
     height 1.5ex depth -.1ex width .2pt\kern-.3em\rm#1}}
 \def\Cx {{\ibb C}} 

\def\QED{{\hspace*{\fill}{\vrule height 1.8ex width 1.8ex }\quad}
    \vskip 0pt plus20pt}

\def\A1n{A_1\otimes\cdots\otimes A_n}

\def\dom{\mathop{\rm dom}\nolimits}

\def\ker{{\rm ker}\,}

\def\Tr{\mathop{\rm Tr}\nolimits}

\def\phi{\varphi}            
\def\epsilon{\varepsilon}    

\def\A{{\cal A}}

\catcode`@=11
\def\@biblabel#1{#1.}
\def\ifundefined#1{\expandafter\ifx\csname
                        \expandafter\eat\string#1\endcsname\relax}
\def\atdef#1{\expandafter\def\csname #1\endcsname}
\def\atedef#1{\expandafter\edef\csname #1\endcsname}
\def\atname#1{\csname #1\endcsname}
\def\ifempty#1{\ifx\@mp#1\@mp}
\def\ifatundef#1#2#3{\expandafter\ifx\csname#1\endcsname\relax
                                  #2\else#3\fi}
\def\eat#1{}
\newdimen\refskip  \refskip=0pt 
\def\@utfirst #1,#2\@ver
   {\author#1,\ifx#2\@ut\afteraut\else\@utsecond#2\@ver\fi}
\def\@utsecond #1,#2\@ver
   {\ifx#2\@ut\andone\author#1,\afterauts\else
      ,\ \author#1,\@utmore#2\@ver\fi}
\def\@utmore #1,#2\@ver
   {\ifx#2\@ut\and\author#1,\afterauts\else
      ,\ \author#1,\@utmore#2\@ver\fi}
\def\authors#1{\@utfirst#1,\@ut\@ver}
\let\more\relax  
\catcode`@=12
\def\Bref#1 "#2"#3{\authors{#1}:\ {\it #2}, #3\more}
\def\Gref#1 "#2"#3{\authors{#1}\ifempty{#2}\else:``#2''\fi, #3\more}
\def\Grefnt#1 "#2"#3{\authors{#1}, #3\more} 
\def\Jref#1 "#2"#3{\relax \authors{#1}:``#2'', \Jn{#3}}
\def\Jrefnt#1 "#2"#3{\relax \authors{#1}, \Jn{#3}} 
\def\inPr#1 "#2"#3\relax{in: \authors{\eds#1}:{\it #2}, #3}
\newcommand{\Jn}[4]{{\it#1}\ {\bf#2} (#3), #4}
\def\author#1. #2,{#2, #1.}
\def\sameauthor#1{\leavevmode$\underline{\hbox to 25pt{}}$}
\def\and{, and }   \def\andone{ and }
\def\noinitial#1{\ignorespaces}
\let\afteraut\relax
\let\afterauts\relax
\def\etal{\def\afteraut{, et.al.}\let\afterauts\afteraut
           \let\and,}
\def\eds{\def\afteraut{(ed.)}\def\afterauts{(eds.)}}
\newcommand{\condmat}[1]{archived as {\tt cond-mat/#1}}
\begin{document}
{\baselineskip=14pt
\thispagestyle{empty}
{Archived as {\tt cond-mat/9512120}\hspace{\fill} Preprint TKBN11-95}
\vspace{30pt}
\begin{center}
{\LARGE\bf The spectral gap of the} \\[10pt]
{\LARGE\bf ferromagnetic XXZ chain} \\[30pt]
Tohru Koma\\
Department of Physics\\
Gakushuin University\\
Mejiro, Toshima-ku\\
Tokyo 171, JAPAN\\
E-mail: {\tt koma@riron.gakushuin.ac.jp}\\[15pt]
Bruno Nachtergaele\\
Department of Physics\\
Princeton University\\
Princeton, NJ 08544-0708, USA\\
E-mail: {\tt bxn@math.princeton.edu}\\[15pt]
(7 November 1995)\\[30pt]

{\bf Abstract}\\[10pt]
\end{center}
We prove that the spectral gap of the spin-1/2 ferromagnetic XXZ chain
with Hamiltonian $H=-\sum_x S^{(1)}_xS^{(1)}_{x+1}+S^{(2)}_xS^{(2)}_{x+1}
+\Delta S^{(3)}_xS^{(3)}_{x+1}$, is given by $\Delta-1$ for all
$\Delta\geq 1$. This is the gap in the spectrum of the infinite chain
in any of its ground states, the translation invariant ones as well as
the kink ground states, which contain an interface between an {\it
up\/} and a {\it down\/} region. In particular, this shows that the
lowest magnon energy is not affected by the presence of a domain wall.
This surprising fact is a consequence of the $SU_q(2)$ quantum group
symmetry of the model.

\vspace{20pt}

\vfill

\hrule width2truein
\smallskip
{\baselineskip=12pt
\noindent
Copyright \copyright\ 1995 by the authors. Faithful reproduction of
this article by any means is permitted for non-commercial purposes.\par
}}
\newpage

\masection{Introduction}\label{sec:intro}

The main purpose of this paper is to determine in a completely
rigorous setting the exact value of the spectral gap of the
spin-1/2 ferromagnetic XXZ Heisenberg chain in the
thermodynamic limit. By the spectral gap we mean the gap above the
ground state in the spectrum of the GNS Hamiltonian in one of the
ground state representations of the model.
The ferromagnetic XXZ chain has translation invariant ground
states as well as ground states that contain a domain wall (the
so-called kink ground states). It is surprising that the gap does not
depend on the reference ground state, i.e., the presence of a
domain wall does not affect the energy of the lowest excited state.
In general the gap in a kink ground
state cannot exceed the gap in the homogeneous ground states (see
Section \ref{sec:upper}). That the two gaps coincide for the
$S=1/2$ XXZ chain is a consequence of the $SU_q(2)$ quantum group
symmetry of the model. The
homogeneous and kink ground states considered in this paper
are {\it all\/} the infinite volume ground states of the XXZ chain
\cite{Mat}.

Our main result is given in Theorem \ref{thm:main} at the end of this
introduction. The expression for the spectral gap given there
coincides with the Bethe Ansatz result
(see e.g.  \cite{JMcC,JKMcC,AKS}). It should be noted that treatments
based on the Bethe Ansatz suffer from the fact that the completeness of
the eigenstates obtained by this method is still an unproven
assumption \cite{KE}.
Our proof does not use the Bethe Ansatz and is free from any
unproven assumptions.

The following finite volume Hamiltonians
that include special boundary terms, will be useful:
\begin{equation}
H^{XXZ}_L=A(\Delta)\left(S_L^{(3)}-S_1^{(3)}\right)
-\sum_{x=1}^{L-1} \left[\Delta^{-1}
\left(S^{(1)}_xS^{(1)}_{x+1}+S^{(2)}_xS^{(2)}_{x+1}\right)
+S^{(3)}_xS^{(3)}_{x+1}\right]
\label{ham+bc}
\end{equation}
with the anisotropic coupling $\Delta\ge 1$,
where $S_x^{(\alpha)} \ (\alpha=1,2,3)$ are the usual $2\times 2$ spin
matrices (with eigenvalues $\pm 1/2$) acting on the site $x$, and
$A(\Delta)=\pm {1\over 2}\sqrt{1-1/\Delta^2}$. For arbitrary finite
volumes $\Lambda$ we denote the corresponding Hamiltonians by
$H^{XXZ}_\Lambda$.

\noindent
The two Hamiltonians
corresponding to the positive and negative choice of $A(\Delta)$ are
obviously unitarily equivalent by left-right symmetry. Unless
explicitly mentioned we will always refer to the Hamiltonian with the
positive choice for $A(\Delta)$. The boundary conditions and
normalization of \eq{ham+bc} are natural for the following reasons.
First of all they make the ground state degeneracy equal to $L+1$ {\it
for all $\Delta\geq 1$}. This property can be explained in terms of a
quantum group symmetry that the Hamiltonian, with these particular
boundary terms included, possesses \cite{PS}. The normalization is
such that one can consider the limit $\Delta\to\infty$ without
difficulty. In this limit the model becomes the ferromagnetic Ising
chain with a boundary term that allows for ground states with a kink,
i.e., for any site $x$ in the finite chain the configuration with all
spins to the left of $x$ {\it up \/} ($\up$) and  all spins to the
right of $x$ {\it down\/} ($\down$), is a ground state. Obviously
there are $L-1$ of such kink states. Together with the two translation
invariant configurations this yields $L+1$ ground states.
As we will see in Section \ref{sec:infinite} the
boundary terms also make the computation of the GNS Hamiltonians
of the infinite chain immediate.

In the isotropic limit ($\Delta=1$) the $L+1$-fold degeneracy is the
dimension of the spin $L/2$ representation of SU(2). Note that the
boundary terms vanish for $\Delta=1$. In the thermodynamic limit
($L\to\infty$) all translation invariant ground states are
states of perfectly aligned spins. No non-translation invariant
ground states are known.

If $\Delta >1$, there are four different classes of ground
states of the model on the infinite chain, which could be called {\it
up, down, kink\/}, and {\it antikink\/}.  They consist, respectively,
of the state with all spins $\up$, the state with all spins $\down$,
an infinite number of states in which the spins are $\up$ at $-\infty$
and $\down$ at $+\infty$, and an infinite number of states in which
the spins are $\down$ at $-\infty$ and $\up$ at $+\infty$. The
infinite degeneracy of the ground state in the latter two sectors
corresponds to the possible choices for the location of the kink or
antikink.
The
kinks are strictly speaking located at a single bond only in the Ising
limit ($\Delta\to\infty$).  For $1<\Delta<\infty$ the ground states
are not described by a single configuration because of the quantum
fluctuations, but the kinks, respectively antikinks, are
quasi-localized for all $1<\Delta<\infty$. All these properties have
been investigated in great detail and with complete mathematical
rigor in a beautiful paper by Gottstein and Werner \cite{GW}. The
kink states themselves were also written down independently by
Alcaraz, Salinas and Wreszinski in \cite{ASW}, who moreover discovered
that the exact expressions of ground states containing a domain wall
generalize directly to higher spin and higher dimensions. The
detailed properties, including the excitation spectrum, of the
interface ground states in two and higher dimensions is the
subject of a separate paper \cite{KN2}.

It is remarkable that the turning over from up to down in a kink
ground state is exponentially localized in this one-dimensional
system. One likes to think of the XXZ ferromagnet as the Ising model
with quantum fluctuations. In low dimensions, phase boundaries, or
domain walls, are unstable against thermal fluctuations. In the XXZ
chain (as well as in its higher-dimensional cousins),
quantum fluctuations do {\it not\/} destroy domain walls as long as
$\Delta>1$.
This can be intuitively understood
as a consequence of the competition between the
$-S^{(1)}_xS^{(1)}_{x+1}-S^{(2)}_xS^{(2)}_{x+1}$ and the
$-S^{(3)}_xS^{(3)}_{x+1}$ terms in the Hamiltonian.  The first term
favors configurations with opposite spins at $x$ and $x+1$. The second
term is attractive and favors parallel spins at neighboring sites.
First, consider the isotropic chain ($\Delta =1$).
There, all ground states have the full permutation symmetry of the
lattice (the set of all states of highest possible spin, $S_{tot}=L/2$,
coincides with the set of states that are invariant under arbitrary
permutations of the lattice sites). So, the ground state at a fixed
value of the third component of the spin is uniquely characterized by
the fact that the positions of the down spins are all equally probable
(and all configurations occur with the same phase). This means that
the two terms in the Hamiltonian exactly balance each other resulting
in ground states that are indifferent for the down spins to be
neighbors or not. It is therefore not so surprising that increasing
the relative weight of the $-S^{(3)}_xS^{(3)}_{x+1}$ term in the
Hamiltonian by any non-vanishing amount, results in an effective
attraction between like spins, leading to phase separation in the
ground state.

The gap in the thermodynamic limit is a well-defined notion only with
respect to one of the aforementioned four superselection
sectors: up, down, kink, and antikink. Each of these sectors
corresponds to a different representation of the observable algebra of
the system. In these representations the Heisenberg dynamics of the
model is generated by a densely defined self-adjoint, non-negative
definite operator $H$. Theorem \ref{thm:main} below refers to the gap
above zero in the spectrum of this operator. Physically it is the gap
usually referred to as the ``bulk gap'', i.e., the gap in the Hilbert
space of states that are quasi-local perturbations of an
infinite volume ground state. In particular, edge excitations are
{\it not} taken into account.

\begin{Thm}\label{thm:main}
For  all $\Delta \geq 1$, and in each of the sectors described
above as {\em up, down, kink\/}, and {\em antikink\/},
the infinite volume gap $\gamma$ is given by
\be
\gamma=1-\Delta^{-1}
\label{gammagap}
\ee
\end{Thm}


\masection{The ground states of the XXZ chain}\label{sec:ground}

Only those aspects of the ground states of the XXZ chain that have
direct relevance to our proof and understanding of the spectral
gap of the model will be presented here. A more detailed  analysis can
be found in \cite{GW}, and various aspect of the ground states have
been discussed in the literature (see e.g. \cite{ASW} and the
references therein).

For the study of the finite chains we use the
special boundary conditions introduced in \eq{ham+bc}.
Up to a constant the Hamiltonian can be written as follows:
\begin{equation}
H^{(q)}_L=H^{XXZ}_L +(L-1)/4=\sum_{x=1}^{L-1}h^q_{x,x+1}
\label{hammu}
\end{equation}
where $h^q_{x,x+1}$ is the orthogonal projection on the vector
\begin{equation}
\xi_q=\frac{1}{1+q^2}(q\ket{\up \down }-\ket{\down \up })
\label{ximu}
\end{equation}
Here the parameter $q$ is given by $\Delta=(q+q^{-1})$ with
the range $0\le q\le 1$.
In terms of the spin matrices $h^q_{1,2}$ is given by
\be
h^q_{1,2}=- \Delta^{-1}(S^{(1)}_1 S^{(1)}_2 +
S^{(2)}_1 S^{(2)}_2) - S^{(3)}_1 S^{(3)}_2
+{1\over 4} +A(\Delta)(S^{(3)}_2 - S^{(3)}_1)
\ee
with $A$ defined following \eq{ham+bc}. From the definition of $\xi_q$
it is obvious that $h^q_{x,x+1}\ket{\up\cdots\up}=0$ for all
$x=1,\ldots,L-1$. As $H^{(q)}_L$ is the sum of the $h^q_{x,x+1}$,
which are positive, this implies that the ground state  energy of
$H^{(q)}_L$ is zero and that $\ket{\up\cdots\up}$ is a ground state.
For all $0\leq q\leq 1$, the ground state space ($\equiv\ker
H^{(q)}_L$) is $L+1$-dimensional. We will often omit the superscripts
$q$ to simplify notations.

For all $\Delta \geq 1$ the uniform states $\ket{\up\cdots\up}$ and
$\ket{\down\cdots\down}$ are ground states of the XXZ chain.
If $\Delta =1$ the $L+1$-dimensional ground state space is the
spin $L/2$ representation of SU(2). For all $\Delta >1$ and
$A=\frac{1}{2}\sqrt{1-1/\Delta^2}$, the non-uniform ground states can
be thought of as kink states, which are roughly described as the Ising
kinks plus quantum fluctuations.
For
$A=-\frac{1}{2}\sqrt{1-1/\Delta^2}$ the kinks have to be replaced by
antikinks, i.e., the roles of $\up$ and $\down$ spins have to be
interchanged (or, equivalently, one can interchange left and right).
We refer to \cite{GW,ASW,KN2} for more details and explicit
expressions.

In the thermodynamic limit the boundary terms disappear to infinity and
the left-right symmetry of the model, broken by the particular
boundary terms we have introduced, must be restored. It is therefore
obvious that both the kink and antikink states appear as infinite
volume ground states of the model.

For our purposes the most convenient way to describe the space of
ground states of a chain of length $L$ is to introduce deformed
raising and lowering operators which, together with the third
component of the spin, generate the algebra (quantum group) of
SU$_q$(2). The representation of SU$_q$(2) on the finite chain is {\it
not\/} left-right symmetric, and is different for the boundary terms
that produce kink and antikink ground states. In fact, the two mutually
non-commuting representations of SU$_q$(2) together generate the
infinite-dimensional quantum affine symmetry algebra
$\widehat{sl_q}(2)$ that lies at the basis of the integrability of the
model (see e.g.  \cite{DFJMN,JM}).  A rigorous formulation of this
infinite dimensional symmetry of the XXZ chain, has not yet been
obtained (see \cite{FQG} for a discussion of some of the problems). We
will not use it here.

Here, we restrict the discussion of the quantum group symmetry of the
XXZ model to the bare minimum (see, e.g., \cite{Kas} for the
representation theory of $SU_q(2)$).  One can think of the quantum
group symmetry as a systematic way to construct operators that commute
with the Hamiltonians $H^{(q)}_L$. The parallelism with the usual
arguments in the ``theory of angular momentum'' in quantum mechanics
(representations of SU(2)) is so perfect that the reader will hardly
notice the difference.

For $0<q<1$ define the $2\times 2$ matrix $t$ by
$t=q^{-2S^{(3)}}$,
and define as usual $S^\pm=S^{(1)}\pm i S^{(2)}$.
It is trivial to check that
$S^\pm$ and $t$ satisfy the following commutation relations
\be
tS^\pm=q^{\mp2}S^\pm t, \mbox{ and \ }
{[ S^+ , S^- ] }=\frac{\displaystyle t-t^{-1}}{\displaystyle
q^{-1}-q} =2S^{(3)}
\label{sumua}
\ee
They are just the SU(2) commutation relations in disguise. The
remarkable fact is that there is a simple definition of the tensor
product (coproduct of the quantum group)
of any two representations of the commutation relations
\eq{sumua}, yielding a new representation. The
representation on a chain of $L$ spins is given by
\alpheqn\bea
S^{(3)}_{[1,L]}&=&\sum_{x=1}^L \idty_1\otimes\cdots\otimes
S^{(3)}_x\otimes\idty_{x+1}\otimes\cdots\idty_L\label{spinmua}\\
S^+_{[1,L]}&=&\sum_{x=1}^L t_1\otimes\cdots\otimes t_{x-1}\otimes
S^+_x\otimes\idty_{x+1}\otimes\cdots\idty_L\label{spinmub}\\
S^-_{[1,L]}&=&\sum_{x=1}^L \idty_1\otimes\cdots\otimes
S^-_x\otimes t^{-1}_{x+1}\otimes\cdots t^{-1}_L\label{spinmuc}
\eea\reseteqn
where we used an index to identify the sites on which the tensor
factors act. Note that, for $L\geq 2$, the operators $S^\pm_{[1,L]}$
depend on $q$ through $t$. One can easily check that the ``spin''
operators as defined in \eq{spinmua}--\eq{spinmuc} commute with the
interaction terms $h^q_{x,x+1}$ and hence with the Hamiltonian
$H^{(q)}_L$ itself. Therefore, all states of the form
$(S^-_{[1,L]})^k\ket{\up\cdots\up}$ are ground states. This way we
obtain $L+1$ ground states. That there are no other ground states
follows from the ground state equations $h^q_{x,x+1}\psi=0$, for
$1\leq x\leq L-1$, and for an arbitrary
$\psi=\sum_{\{\sigma_x=\up,\down\}}\psi(\{\sigma_x\})\ket{\{\sigma_x\}}$.
This set of equations is equivalent to the equations
$\psi(\cdots\down\up\cdots) =q \psi(\cdots\up\down\cdots)$, where the
$\up$ and $\down$ spins are at the sites $\{x,x+1\}$.  It is then
obvious that there is only one solution in each sector of fixed number
of $\down$ spins.


\masection{The gap for finite chains}\label{sec:gap}

In the proof of Theorem \ref{thm:main}
we will need the exact spectral gap for finite chains with the boundary
conditions of \eq{ham+bc}. As will become clear in Section
\ref{sec:infinite}, this choice of boundary conditions is the optimal
one and essentially the only choice that when taking the thermodynamic
limit will yield a lower bound that is, in fact, the exact gap in the
thermodynamic limit.

Denote by $\gamma_L$ the spectral gap of $H^{(q)}_L$, which, as the
ground state energy vanishes,  is equal to the energy of first excited
state and let $\epsilon_L$ be the first excited state in the sector
with exactly one down spin. The aim of this section is to prove the
following proposition.

\begin{Prop}\label{prop:finitevolume}
For the $SU_q(2)$ invariant
spin-1/2 ferromagnetic XXZ chain with Hamiltonian
\eq{ham+bc}, $L\geq 2$, and $\Delta \geq 1$ one has
\be
\gamma_L=\epsilon_L=1-\Delta^{-1}\cos(\pi/L)
\label{gammaL}\ee
and in particular
\be
\gamma_L\geq 1-\Delta^{-1}\quad .
\label{gamma}\ee
\end{Prop}

The proof of this proposition combines two rather elementary facts:
\begin{description}
\item{1)} There is a first excited
state of the chain of $L$ sites with total spin equal to $(L/2) -1$,
i.e., the maximal possible value of the total spin minus one (see
Lemma \ref{lem:onedown}).
\item{2)} In the sector of total spin $(L/2) -1$, $H_L^{(q)}$
can easily be diagonalized (see Lemma \ref{lem:epsilonn}).
\end{description}

The remainder of this section is devoted to proving two lemmas that
together establish 1) and 2) and hence prove \eq{gammaL}.

Consider an arbitrary spin chain of $L$ sites and with Hilbert space
$\Hs_L=\bigotimes_{x=1}^L (\Cx^d)_x$,
and with local Hamiltonians of the following form:
\be
H_L=\sum_{x=1}^{L-1} h_{x,x+1}
\label{HL}
\ee
where $h_{x,x+1}$ acts non-trivially
only at the nearest neighbor pair $\{x,x+1\}$ and $h_{x,x+1}\geq 0$.
Assume that $\ker H_L \neq\{0\}$.
It is obvious that $\ker H_L =\bigcap_{x=1}^{L-1} \ker h_{x,x+1}$.
For an arbitrary subset
$\Lambda$ let $G_\Lambda$ be the orthogonal projection onto
\be
\ker \sum_{x, \{x,x+1\}\subset\Lambda} h_{x,x+1}
\label{GLambda}
\ee
For intervals $[a,b]$, $1\leq a < b\leq L$, $G_{[a,b]}$ is the
orthogonal projection onto the zero eigenvectors of $\sum_{x=a}^{b-1}
h_{x,x+1}$, and $G_{\{x\}}=\idty$ for all $x$. It then follows
that
\alpheqn\bea
G_{\Lambda_2}G_{\Lambda_1} &=& G_{\Lambda_1}G_{\Lambda_2}
= G_{\Lambda_2} \mbox{ if } \Lambda_1 \subset\Lambda_2
\label{GLambdaa}\\
G_{\Lambda_1}G_{\Lambda_2} &=& G_{\Lambda_2}G_{\Lambda_1}
\mbox{ if } \Lambda_1 \cap \Lambda_2 =\emptyset
\label{GLambdab}
\eea\reseteqn
We will often write $G_n$ instead of $G_{[1,n]}$.
Define operators $E_n$, $1\leq n\leq L$, on $\Hs_L$ by
\be
E_n=\cases{\idty-G_{[1,2]}       & if $n=1$\cr
           G_{[1,n]}-G_{[1,n+1]} & if $2\leq n \leq L-1$\cr
            G_{[1,L]}            & if $n=L$\cr }
\label{defEn}\ee
One can then easily verify, using the properties
\eq{GLambdaa}-\eq{GLambdab}, that $\{E_n \mid1\leq n\leq L\}$
is a family of mutually orthogonal projections summing up to $\idty$.

Next, we add the assumption that the interaction terms are $SU(2)$
or $SU_q(2)$ invariant, and that the space of all ground states
for a finite chain is the irreducible representation of maximal
spin. Of course, $SU(2)$ is a special case of $SU_q(2)$
($q=1$). We treat the cases of $SU_q(2)$ and $SU(2)$ in exactly
the same way and will therefore not distinguish between them.
E.g., we will label the irreducible representations by their
dimension $2s+1$ using a half-integer $s$ that we will call
``spin'' both in the group and the quantum group case.
There is still no need to restrict ourselves to spin-1/2 chains.
So, let $H_n$ be a Hamiltonian for a spin-$s$ chain of $n$ sites,
with arbitrary $s=1/2,1,3/2,\ldots$.
Define $\epsilon_n^{(J)}$ by
\be
\epsilon_n^{(J)}=\mathop{\min_{0\neq\psi\perp {\rm ker}H_n,}}_
{\psi \in \Hs_{S^{(3)}\geq ns-J}}
\frac{\langle\psi\vert H_n \psi\rangle}{\Vert\psi\Vert^2}
\label{epsilonJ}\ee
where $\Hs_{S^{(3)}\geq ns-J}$ is the subspace of $\Hs_{[1,n]}$
where $S^{(3)}_{[1,n]}\geq ns-J$, i.e., the subspace spanned by the
eigenvectors of $S^{(3)}_{[1,n]}$ with eigenvalues
$=ns, ns-1,\ldots,ns-J$.

\begin{Lem}\label{lem:onedown}
Consider an $SU_q(2)$ invariant spin-$s$ ferromagnetic chain of
$L$ sites with a nearest neighbor Hamiltonian
$H_L=\sum_{x=1}^{L-1} h_{x,x+1}$, and for which the  space of all
ground states of a finite chain of $n$ sites is the irreducible
representation of maximal spin ($=ns$), for $2\leq n\leq L$.
Let $\gamma_n$ denote the spectral gap of $H_n$ and let
$\epsilon_n^{(J)}$ be defined by \eq{epsilonJ}.
If
\be
\epsilon_n^{(2s)}\ge \epsilon_{n+1}^{(2s)}
\label{strict}\ee
for all $n, 2\leq n\leq L-1$, then
\be
\gamma_L=\epsilon_L^{(2s)}
\ee
\end{Lem}
\begin{proof}
We will first show that for all $n, 2\leq n\leq L-1$
at least one of the following must be true:
{\it i)\/} $\gamma_{n+1}= \epsilon_{n+1}^{(2s)}$, or
{\it ii)\/} $\gamma_{n+1}\geq \gamma_n$.

Let $\phi_{n+1}$ be an eigenvector of $H_{n+1}$ with eigenvalue
$\gamma_{n+1}$. We can assume that $\phi_{n+1}$ is also an eigenvector
of the Casimir operator (i.e., ${\bf S}_{[1,n+1]}^2$,
in the $SU(2)$ case). This means that $\phi_{n+1}$ belongs to an
irreducible representation.
We distinguish two cases:
{\it a)\/} $E_n\phi_{n+1}\neq 0$, and
{\it b)\/} $E_n\phi_{n+1}= 0$.
Here $E_n$ is the projection defined in \eq{defEn}.
We show that a) implies i) and b) implies ii), and therefore,
i) or ii) (or both) must hold.

{\it case a):\/} If $E_n\phi_{n+1}\neq 0$ there is a $\psi$ in the
range of the projection $E_n$ such that
$\langle\psi\mid\phi_{n+1}\rangle \neq 0$. $E_n\psi=\psi$ is
equivalent to $G_n\psi=\psi$ and $G_{n+1}\psi=0$.
$G_{[1,n]}\psi=\psi$ implies that $\psi\in D^{(ns)}_{[1,n]}\otimes
D^{(s)}_{\{n+1\}}\subset \Hs_{n+1}$, where $D^{(J)}_\Lambda$ denotes a
spin $J$ representation of $SU_q(2)$ acting on $\Hs_\Lambda$.  As
$D^{(ns)}\otimes D^{(s)}\cong D^{((n-1)s)}\oplus \cdots \oplus
D^{((n+1)s)}$, $G_{[1,n+1]}\psi=0$ implies that $\psi\in
\bigoplus_{J=1}^{2s} D^{((n+1)s-J)}$.  Because, by assumption,
$\phi_{n+1}$ belongs to an irreducible representation, the fact that
it is not orthogonal to $\psi$, implies that $\phi_{n+1}$
belongs to an irreducible representation $D^{(J)}$ with
$J\in\{(n+1)s-1,(n+1)s-2,\ldots,(n-1)s\}$. From the definition of
$\epsilon_{n+1}^{(2s)}$ in \eq{epsilonJ} it is then obvious that
$\gamma_{n+1}=\epsilon_{n+1}^{(2s)}$.

{\it case b):\/} If $E_n\phi_{n+1}= 0$ we have also $G_n\phi_{n+1}=0$
because by assumption $G_{n+1}\phi_{n+1}=0$. But then
\be
\gamma_{n+1} =
\frac{\langle\phi_{n+1}\vert H_{n+1}\phi_{n+1}\rangle}{\Vert \phi_{n+1}
\Vert^2}\geq
\frac{\langle\phi_{n+1}\vert H_n\phi_{n+1}\rangle}{\Vert \phi_{n+1}
\Vert^2}=\frac{\langle\phi_{n+1}\vert(\idty-G_n)H_n(\idty-G_n)
\phi_{n+1}\rangle}{\Vert (\idty-G_n)\phi_{n+1}\Vert^2}
\geq\gamma_n
\ee
Here, both $G_n$ and $H_n$ are considered as operators on
$\Hs_{n+1}$, and we used the obvious bound
$(\idty-G_n) H_n(\idty-G_n)\geq \gamma_n (\idty-G_n)$.

The proof of the lemma can now be completed by contraposition.
$\gamma_2=\epsilon_2^{(2s)}$ is obvious.  Let $m>2$ be the smallest
integer for which $\gamma_m\neq \epsilon_m^{(2s)}$. As $\gamma_n \leq
\epsilon_n^{(2s)}$, for all $n$, this means $\gamma_m <
\epsilon_m^{(2s)}$.  Therefore, i) from above cannot hold with
$n+1=m$, and hence ii) must hold, i.e., $\gamma_m\geq\gamma_{m-1}$.
We assumed that $m$ was the {\it smallest\/} integer such that
$\gamma_m \neq \epsilon_m^{(2s)}$, hence $\gamma_{m-1}
=\epsilon_{m-1}^{(2s)}$. We conclude that $\epsilon_m^{(2s)} >
\gamma_m\geq \gamma_{m-1} =\epsilon_{m-1}^{(2s)}$, which is in
contradiction with the assumption \eq{strict}. Therefore, no $m \geq
2$ such that $\gamma_m\neq \epsilon_m^{(2s)}$ exists and the lemma is
proved.
\end{proof}

We now return to the spin-1/2 XXZ Heisenberg chain for the
computation of $\epsilon_n^{(1)}$.

\begin{Lem}\label{lem:epsilonn}
For the $SU_q(2)$ invariant
spin-1/2 ferromagnetic XXZ chain with $\Delta \geq 1$,
$\epsilon_n^{(1)}$ defined in \eq{epsilonJ} is given by
\be
\epsilon_n^{(1)}=1-\Delta^{-1}\cos (\pi/n)
\ee
In particular one has $\epsilon_{n+1}^{(1)} < \epsilon_n^{(1)}$.
\end{Lem}
\begin{proof}
We calculate the eigenvalues ${\cal E}$ of the Hamiltonian
$H^{(q)}_L$ of (\ref{hammu}) in the one down spin sector
by using a transfer matrix method.
An arbitrary vector $\psi$ in that subspace can be written as
\begin{equation}
\psi=\sum_{x=1}^L a_x D_x ,
\end{equation}
where $D_x$ denotes the basis vector with all spins up except
at the site $x$ where the spin is down.
For $\psi$ to be an eigenvector the coefficients $a_x$ must satisfy
$\langle D_y|H^q_L\psi\rangle={\cal E} a_y$
where ${\cal E}$ is the eigenvalue, which amounts to
\begin{equation}
a_{y+1}=2\Delta(1 -{\cal E})a_y-a_{y-1} \quad
\mbox{for \quad $2\le y \le L-1$},
\label{naka}
\end{equation}
\begin{equation}
a_2=2\Delta[1/2+A(\Delta)-{\cal E}]a_1,\quad
a_{L-1}=2\Delta[1/2-A(\Delta)-{\cal E}]a_L.
\label{edge1}
\end{equation}
The equations (\ref{naka}) can be rewritten as
\begin{equation}
\left(
\begin{array}{c}
a_{y+1}\\
a_y
\end{array}
\right)=T
\left(
\begin{array}{c}
a_y \\
a_{y-1}
\end{array}
\right)\quad,\quad \mbox{with} \quad
T=
\left(
\begin{array}{cc}
2\Delta(1-{\cal E})& -1\\
1&0
\end{array}
\right).
\label{transfer}
\end{equation}
By using (\ref{transfer}) repeatedly, we get
\begin{equation}
\left(
\begin{array}{c}
a_L\\
a_{L-1}
\end{array}
\right)=T^{L-2}
\left(
\begin{array}{c}
a_2\\
a_1
\end{array}
\right).
\end{equation}
Combining this with (\ref{edge1}),
we have
\begin{equation}
a_L
\left(
\begin{array}{c}
1\\
2\Delta{[1/2-A(\Delta)-{\cal E}]}
\end{array}
\right)=a_1T^{L-2}
\left(
\begin{array}{c}
2\Delta{[1/2+A(\Delta)-{\cal E}]}\\
1
\end{array}
\right).
\label{Teq}
\end{equation}

This equation can be solved in terms of the eigenvalues and -vectors
of the transfer matrix $T$. 
The eigenvalues $\lambda$ of $T$ are given by the roots of
the equation
\begin{equation}
\lambda^2-2\Delta(1-{\cal E})\lambda+1=0,
\label{chara}
\end{equation}
given by
\begin{equation}
\lambda_{\pm}=\Delta(1-{\cal E})\pm
\sqrt{\Delta^2(1-{\cal E})^2-1}.
\label{lam}
\end{equation}

Consider first the case $\Delta(1-{\cal E})\ne \pm1$.
Then the eigenvectors are determined by
\begin{equation}
u_\pm=
\left(
\begin{array}{c}
\lambda_\pm\\
1
\end{array}
\right).
\end{equation}
In terms of 
$u_\pm$, the vector with $A(\Delta)=\sqrt{1-\Delta^{-2}}/2$
in (\ref{Teq}) can be rewritten as
\begin{equation}
\left(
\begin{array}{c}
2\Delta{[1/2+A(\Delta)-{\cal E}]}\\
1
\end{array}
\right)
=
\left(
\begin{array}{c}
\Delta+\sqrt{\Delta^2-1}-2\Delta{\cal E} \\
1
\end{array}
\right)=\alpha_+ u_+ +\alpha_- u_-
\label{vec1}
\end{equation}
with
\begin{equation}
\alpha_\pm =
{1 \over 2}\left[1\pm
{1 \over \sqrt{\Delta^2(1-{\cal E})^2-1}}
(\sqrt{\Delta^2-1}-\Delta{\cal E})\right] .
\label{alpha}
\end{equation}
Similarly the vector in the left-hand side of (\ref{Teq})
can be rewritten as
\begin{equation}
\left(
\begin{array}{c}
1\\
\Delta-\sqrt{\Delta^2-1}-2\Delta{\cal E}
\end{array}
\right)
=\beta_+
\left(
\begin{array}{c}
1\\
\lambda_-
\end{array}
\right)
+\beta_-
\left(
\begin{array}{c}
1\\
\lambda_+
\end{array}
\right)
=\beta_+\lambda_- u_+ +\beta_-\lambda_+ u_-
\label{vec2}
\end{equation}
with
\begin{equation}
\beta_\pm =
{1 \over 2}\left[1\pm
{1 \over \sqrt{\Delta^2(1-{\cal E})^2-1}}
(\sqrt{\Delta^2-1}+\Delta{\cal E})\right] ,
\label{beta}
\end{equation}
where we have used $\lambda_+\lambda_-=1$.
Substituting (\ref{vec1}) and (\ref{vec2}) into (\ref{Teq}),
we have
\begin{equation}
({a_L/a_1})\left[\beta_+\lambda_-u_++
\beta_-\lambda_+u_-\right]
= T^{L-2}\left(\alpha_+ u_++\alpha_- u_-\right)
=\alpha_+\lambda_+^{L-2} u_+
+\alpha_-\lambda_-^{L-2} u_-.
\end{equation}
Here we have assumed $a_1\ne 0$. Actually $a_1=0$ implies $\psi=0$.
Since the vectors $u_+$ and $u_-$ are
independent of each other, we get
\begin{equation}
(a_L/a_1)\beta_+=\alpha_+\lambda_+^{L-1}
\label{lam+}\quad \mbox{ and}\quad
(a_L/a_1)\beta_-=\alpha_-\lambda_-^{L-1}.
\end{equation}

If $\alpha_-=0$, we get ${\cal E}=0$, $\beta_-=0$,
$\alpha_+=\beta_+=1$ and $\lambda_+=\Delta+\sqrt{\Delta^2-1}$ from
(\ref{alpha}) and (\ref{beta}). 
The eigenvalue ${\cal E}=0$ is 
the ground state in the one down spin sector.

When $\alpha_-\ne 0$,
$\alpha_\pm,\beta_\pm$ are all non-vanishing. Therefore,
from (\ref{lam+}), we have
\begin{equation}
\lambda_+^{2L-2}=
{\alpha_- \over \alpha_+}\times {\beta_+ \over \beta_-} ,
\label{lam+2}
\end{equation}
where we have used $\lambda_+\lambda_-=1$.
Note that
\begin{equation}
{\alpha_- \over \alpha_+}
={\Delta -\sqrt{\Delta^2-1}\over \lambda_+}\times
{\lambda_+-(\Delta+\sqrt{\Delta^2-1}) \over
\lambda_+-(\Delta-\sqrt{\Delta^2-1})}
\end{equation}
and
\begin{equation}
{\beta_+ \over \beta_-}
={\Delta +\sqrt{\Delta^2-1}\over \lambda_+}\times
{\lambda_+-(\Delta-\sqrt{\Delta^2-1}) \over
\lambda_+-(\Delta+\sqrt{\Delta^2-1})}
\end{equation}
from (\ref{alpha}), (\ref{beta}) and (\ref{lam}).
Combining these with the above (\ref{lam+2}), we have
$\lambda_+^{2L}=1$.
This implies that $\lambda_+=e^{i\pi \ell /L}$,
with $\ell$ an integer. From (\ref{lam}),
we get the energy eigenvalues
\begin{equation}
{\cal E}_L(\ell)=1-{\lambda_++\lambda_-\over 2\Delta}
=1-\Delta^{-1}\cos(\pi \ell /L).
\label{sol}
\end{equation}
Here $\ell =1,2,\ldots,L-1$ because $\lambda_\pm\ne \pm 1$ which
are the degenerate roots of (\ref{chara})
when $\Delta(1-{\cal E})=\pm 1$.
%
Since we have found $L$ distinct eigenvalues, we obtained the complete
set of eigenvalues.
In particular this implies that there are no solutions with
$\Delta(1-{\cal E})=\pm 1$ (except when $\Delta =1$).
\end{proof}


\masection{The infinite chain}\label{sec:infinite}

Before we can prove rigorous statements about the spectrum of
the infinite chain we need to introduce the mathematical objects that
define the infinite system. Although all interesting properties of the
infinite chain can be expressed as results for limits of quantities
defined for finite chains, the converse is not true. Not all limits of
finite chain quantities give interesting or even sensible statements
about the infinite chain.

Let the symbols $\up\up,\up\down,\down\up,\down\down$ denote the four
superselection sectors of the infinite XXZ chain with $\Delta>1$,
corresponding to {\it up, kinks, antikinks\/}, and {\it down\/}
respectively. We can describe the GNS Hilbert spaces \cite{BraRo} of
these four superselection sectors as the so-called incomplete tensor
products \cite{Gui} $\Hs_{\alpha\beta}$, for $\alpha$ and $\beta=\up$
or $\down$, defined by
\begin{equation}
\Hs_{\alpha\beta}=\overline{\bigcup_\Lambda
\left(\bigotimes_{x\in\Lambda}
\Cx^2\otimes\bigotimes_{y\in\Lambda^c}\Omega_{\alpha\beta}(y)\right)}
\label{Hab}
\end{equation}
where
\begin{equation}
\Omega_{\alpha\beta}(y)=
\cases{\ket{\alpha}&if $y\leq 0$\cr
       \ket{\beta}&if $y> 0$\cr}
\label{Oaby}
\end{equation}
We also define the vectors $\Omega_{\alpha\beta}$ as the infinite
product vectors
\begin{equation}
\Omega_{\alpha\beta}=\bigotimes_{y\in\Ir}\Omega_{\alpha\beta}(y)
\in\Hs_{\alpha\beta}
\label{Oab}
\end{equation}
Let $\A_\Lambda$ denote the local observables acting non-trivially only
on the sites in the finite set $\Lambda$. Local observables
$X\in\A_\Lambda$ act on $\Hs_{\alpha\beta}$ in the obvious way, e.g.,
the spin matrices at the site $x$ act on the $x^{th}$ factor of the
tensor product \eq{Hab}. From the definitions above it is clear that
vectors $\psi$ of the form
\begin{equation}
\psi=X\Omega_{\alpha\beta},\qquad X\in \bigcup_\Lambda \A_\Lambda
\label{psi}
\end{equation}
form a dense subspace of $\Hs_{\alpha\beta}$. Note that if
$\alpha\neq\beta$, $\Omega_{\alpha\beta}$ is {\it not\/} the GNS
vector representing one of the kink (or antikink) ground states.
Let $\Omega^{GNS}_{\alpha\beta}$ denote the GNS vector in the
$\alpha\beta$ sector, or one of the GNS vectors in the case of kinks or
antikinks. That $\Omega^{GNS}_{\alpha\beta}\in\Hs_{\alpha\beta}$
follows from the explicit expansion \cite{GW}
$$
\Omega^{GNS}_{\alpha\beta}=Z(q)^{-1}\sum_{k=0}^{\infty}
\sum_{x_1<x_2<\cdots<x_k\leq 0 <y_1<\cdots <y_k}
q^{\sum_{j=1}^k (y_j-x_j)}\prod_{j=1}^k
\sigma^{(1)}_{x_j}\sigma^{(1)}_{y_j}\Omega_{\alpha\beta}
$$
where $\sigma^{(1)}=2S^{(1)}$,
and $Z(q)$ is the normalization factor given by
$$
Z(q)^2=\sum_{k=0}^{\infty}
\sum_{x_1<x_2<\cdots<x_k\leq 0 <y_1<\cdots <y_k}
q^{2\sum_{j=1}^k (y_j-x_j)} <+\infty
$$

The Hamiltonian
is represented on $\Hs_{\alpha\beta}$ as the generator
$H_{\alpha\beta}$ of the Heisenberg dynamics of observables acting on
$\Hs_{\alpha\beta}$.  The dense subspace of the vectors $\psi$ defined
in \eq{psi} is in the domain of $H_{\alpha\beta}$, and the selfadjoint
operator $H_{\alpha\beta}$ is uniquely determined by the requirement
\begin{equation}
H_{\alpha\beta}X\Omega^{GNS}_{\alpha\beta}
=\lim_{\Lambda\to\Ir}[H^{XXZ}_\Lambda,X]\Omega^{GNS}_{\alpha\beta}
\label{hamab}
\end{equation}

We remark that $H_{\alpha\beta}$ does not depend on boundary terms such
as $A(S^{(3)}_a-S^{(3)}_b)$ added to
the XXZ Hamiltonian for finite chains. It
is well-known \cite{BraRo} that $H_{\alpha\beta}$ is a positive
operator in any ground state representation and in the present case
this could not be more clear.
An explicit formula for $H_{\alpha\beta}$ is
\begin{equation}
H_{\alpha\beta}X\Omega_{\alpha\beta}= \sum_{\{x,x+1\}\cap (\Lambda
\cup\{0\})
\neq \emptyset} h^{\alpha\beta}_{x,x+1} X\Omega_{\alpha\beta}
\label{hamloc}
\end{equation}
for arbitrary $X\in {\cal A}_\Lambda$, and
where $h^{\alpha\beta}_{x,x+1}$ can be taken to be $h^q_{x,x+1}$ if
$\alpha\beta=\up\up, \down\down$, or $\up\down$. If $\alpha\beta
=\down\up$ the sign of the boundary term has to be reversed.

The spectral gap $\gamma_{\alpha\beta}$ is then just the gap
above $0$ in the spectrum of $H_{\alpha\beta}$, i.e.,
\be
\gamma_{\alpha\beta}=\inf_{0\neq\psi\perp\ker
H_{\alpha\beta}\atop \psi\in\dom H_{\alpha\beta}}\frac{\langle\psi\mid
H_{\alpha\beta}\psi\rangle}{\langle\psi\mid\psi\rangle}
\label{defGamma}
\ee
There is no a priori reason why the spectrum,
of $H_{\alpha\beta}$ should be independent of the
reference ground state.
We already know that the multiplicity of the lowest
eigenvalue is different: it is 1 for $H_{\up\up}$ and $H_{\down\down}$
and infinite for $H_{\up\down}$ and $H_{\down\up}$. Therefore, a
priori, we should not expect $\gamma_{\alpha\beta}$ to be independent
of $\alpha\beta$. One can easily convince oneself, however, that
$\gamma_{\alpha\beta}= \gamma_{\beta\alpha}$ and that
$\gamma_{\up\up}=\gamma_{\down\down}$. From a simple argument given in
Section \ref{sec:upper} it follows that
$\gamma_{\alpha\beta}\leq\gamma_{\up\up}$. The upper and lower bounds
that we derive here {\it are\/} independent of
$\alpha\beta$, and they are equal, thus proving Theorem \ref{thm:main}.

\masubsection{Proof that $1-\Delta^{-1}$ is a lower bound}
\label{sec:lower}

In order to prove that $1-\Delta^{-1}$ is a lower bound of the gap we
simply have to show that the lower bound \eq{gamma} on the finite
volume gap obtained in Section \ref{sec:gap} remains valid in the
thermodynamic limit, irrespective of the particular zero energy ground
state that we are considering. It is important that the finite volume
gap estimates were obtained for the ``correct'' boundary conditions of
\eq{ham+bc}.  More explicitly we show that for any choice of
$\alpha\beta$ and all local observables $X$ the following inequality
holds:
\be
\langle\Omega_{\alpha\beta}\mid X^* H_{\alpha\beta}^3 X
\Omega_{\alpha\beta}\rangle
\geq (1-\Delta^{-1})\langle\Omega_{\alpha\beta}\mid X^*
H_{\alpha\beta}^2 X\Omega_{\alpha\beta}\rangle
\label{lowerbound}
\ee
This proves that $1-\Delta^{-1}$ is a lower bound for the gap
because the vectors of the form \eq{psi} are a core for all
powers of $H_{\alpha\beta}$.

The inequality \eq{lowerbound} follows from Proposition
\ref{prop:finitevolume} when one observes that for $X\in\A_\Lambda$
\be
\langle\Omega_{\alpha\beta}\mid X^* H_{\alpha\beta}^3 X
\Omega_{\alpha\beta}\rangle
= \langle\Omega_{\alpha\beta}\mid X^* (H^{(q)}_{\Lambda\pm 3})^3 X
\Omega_{\alpha\beta}\rangle
\label{local}
\ee
Obviously, $X^* (H^{(q)}_{\Lambda\pm 3})^3 X\in \A_{\Lambda\pm 3}$.
Therefore the expectation value in the right side of \eq{local} can be
computed in the density matrix $\rho_{\Lambda\pm 3}$ which describes
the state $\Omega_{\alpha\beta}$ in the finite volume $\Lambda \pm 3$.
The same is true for the right side of \eq{lowerbound}. We conclude
that it is sufficient to ascertain that
\be
\Tr \rho_{\Lambda\pm 3} X^* (H^{(q)}_{\Lambda\pm 3})^3 X
\geq (1-\Delta^{-1})\Tr \rho_{\Lambda\pm 3} X^* (H^{(q)}_{\Lambda
\pm 3})^2 X
\ee
which immediately follows from the finite volume result of
Proposition \ref{prop:finitevolume}.
\endproof

\masubsection{Proof that $1-\Delta^{-1}$ is an upper bound}
\label{sec:upper}

First we argue that it suffices to prove the upper bound for
$H_{\up\up}$. It is obvious that the gap of $H_{\down\down}$ will
satisfy the same bound. For the gap of the model in the kink and
antikink sectors we have an inequality which can be derived as
follows. The translation invariant ground states can be  obtained as
weak limits of the kink or antikink states by letting the position of
the kink (or antikink) tend to $\pm\infty$. We then have
\bea
\lefteqn{\inf_{\Lambda,X\in\A_\Lambda}
\frac{\langle\Omega_{\alpha\beta}\mid X^* (H^{(q)}_{\Lambda\pm 3})^3 X
\Omega_{\alpha\beta}\rangle}{\langle\Omega_{\alpha\beta}\mid X^*
(H^{(q)}_{\Lambda\pm 3})^2 X \Omega_{\alpha\beta}\rangle}}\\
&\leq \displaystyle \inf_{\Lambda,X\in\A_\Lambda} \lim_{n\to\pm\infty}
\frac{\langle\Omega_{\alpha\beta}\mid\tau_n(X^*
(H^{(q)}_{\Lambda\pm 3})^3 X)
\Omega_{\alpha\beta}\rangle}{\langle\Omega_{\alpha\beta}\mid \tau_n(X^*
(H^{(q)}_{\Lambda\pm 3})^2 X)\Omega_{\alpha\beta}\rangle}
= \displaystyle\inf_{\Lambda,X\in\A_\Lambda}
\frac{\langle\Omega_{\up\up}\mid X^* (H^{(q)}_{\Lambda\pm 3})^3 X
\Omega_{\up\up}\rangle}{\langle\Omega_{\up\up}\mid X^*
(H^{(q)}_{\Lambda\pm 3})^2 X \Omega_{\up\up}\rangle}
\label{compare}\eea
where $\tau_n$ denotes the translation over $n$ lattice units
in the chain. It follows that $\gamma_{\alpha\beta}\leq
\gamma_{\up\up}$.

For the proof of the upper bound on $\gamma_{\up\up}$
we use the variational principle \eq{defGamma}
and observe that the spaces
$\ker H^{(q)}_\Lambda\subset\Hs_{\alpha\beta}$ are decreasing
in $\Lambda$. Therefore, in order to assure that
$\psi\perp\ker H_{\alpha\beta}$, it suffices to check that
$\psi\perp\ker H^{(q)}_\Lambda$ for some suitable $\Lambda$.

Fix an interval $[1,n]$ and introduce the usual spin wave operators
$X_k$, $k=2\pi m/n, m=0,1,\ldots, n-1$, given by
\be
X_k=\frac{1}{\sqrt{n}}\sum_{x=1}^n e^{ikx} S^-_x
\label{Xk}
\ee
The normalization and the allowed values for $k$ are chosen such that
\be
\langle\Omega_{\up\up}\mid X_l^* X_k\Omega_{\up\up}\rangle=\delta_{k,l}
\label{ortho}
\ee
The vectors $\psi$ we need for the upper bound are linear combinations
of two spin waves, i.e. $\psi=(c_1 X_{k_1}+c_2
X_{k_2})\Omega_{\up\up}$. Due to \eq{ortho} we have $\Vert
\psi\Vert^2=\vert c_1\vert^2 + \vert c_2\vert^2$. For any pair of
distinct $k_1,k_2$, the coefficients $c_1, c_2$ can be chosen such
that $G_{[1,n]}\psi=0$, i.e. $\psi \perp \ker H^{(q)}_{[1,n]}$. This
follows from the fact that $\ker H^{(q)}_{[1,n]}$ contains exactly one
vector for each eigenvalue of $S^{(3)}_{[1,n]}$. All vectors
$X_k\Omega_{\up\up}$ have $S^{(3)}_{[1,n]}=(n-2)/2$.
It follows that any two-dimensional space of vectors $\psi$
with fixed, distinct $k_1,
k_2$ and arbitrary $c_1,c_2$ must contain a ray $\perp \ker
H^{(q)}_{[1,n]}$. Hence the upper bound can be proved by showing that
\be
\inf_{\mathstrut n,k_1,k_2}\sup_{\mathstrut c_1,c_2}
\frac{\langle \psi \mid H_{\up\up}\psi\rangle}{\langle
\psi\mid\psi\rangle} = 1-\Delta^{-1}
\label{upperbound}
\ee
which we do next. From the definition \eq{Xk} of the $X_k$
it is clear that the only matrix elements of $H_{\up\up}$ we
need are the $T_{x,y}$, $1\leq x,y\leq n$, defined by
\be
T_{x,y}=\langle\Omega_{\up\up}\mid S^+_x H^{(q)}_{[0,n+1]}
S^-_y\Omega_{\up\up}\rangle
=\frac{1}{2\Delta}\left\{2\Delta\delta_{x,y}-\delta_{x,y-1}
-\delta_{x,y+1}\right\}
\label{Txy}
\ee
It is then easily seen that the $\sup_{c_1,c_2}$ in the left side of
\eq{upperbound} yields the norm of the $2\times 2$ matrix
$M(n,k_1,k_2)$ with matrix elements $M(n,k_1,k_2)_{i,j}=M_n(k_i,k_j)$
where $M_n(k,l)$, for $k,l$ of the form $2\pi m/n$, is the function
\be
M_n(k,l)=\frac{1}{n}\sum_{x,y=1}^n e^{-ikx}T_{x,y}e^{ily}
        =\delta_{k,l}(1- \Delta^{-1}\cos k)
            + (e^{il}+e^{-ik})/(2\Delta n)
\label{mkl}\ee
It is now obvious that $\inf_{n,k_1,k_2}\Vert M(n,k_1,k_2)\Vert
= 1 -\Delta^{-1}$.
\endproof

\vspace{10pt}\noindent
{\large\bf Acknowledgements}

It is a pleasure to thank the following people for discussions
and correspondence: G.~Albertini, J.~Gr\"{u}neberg,
A.~Kl\"{u}mper, V.~Korepin, E.H.~Lieb, K.~Schoutens,
\hbox{J.-Ph.}~Solovej, H.~Tasaki, and R.F.~Werner. We are  grateful to
R.F.~Werner and W.F.~Wre\-szinski for making their work (\cite{GW} and
\cite{ASW}) available to us prior to publication. This manuscript was
finished during a stay of the authors at the Erwin Schr\"odinger
Institute (Vienna).  B.N. is partially supported by the U.S.  National
Science Foundation under Grant No. PHY90-19433 A04.

\def\thebibliography#1{
    \vspace{10pt}\noindent
    {\large\bf References} \par \list
  {\arabic{enumi}.}{\settowidth\labelwidth{[#1]}
    \leftmargin\labelwidth
    \advance\leftmargin\labelsep
    \itemsep=0pt
    \usecounter{enumi}}
    \def\newblock{\hskip .11em plus .33em minus -.07em}
    \sloppy
    \sfcode`\.=1000\relax}

{\renewcommand{\baselinestretch}{1.2}\large
}
\end{document}
\bye